\documentclass[aps,prl,reprint,groupedaddress,showpacs]{revtex4-1}
\usepackage{amssymb}
\usepackage{graphicx}
\usepackage{hyperref}
\usepackage[bottom]{footmisc}
\usepackage[switch, modulo]{lineno}

\begin{document}

\title{Fractal Dimension of Particle Showers Measured in a Highly Granular Calorimeter}


\author{Manqi Ruan$^{1, 2}$}
\email[Corresponding author:]{ Manqi.Ruan@ihep.ac.cn}
\author{Daniel Jeans$^{1, 3}$}
\author{Vincent Boudry$^{1}$}
\author{Jean-Claude Brient$^{1}$}
\author{Henri Videau$^{1}$}

\affiliation{$^1$Laboratoire Leprince-Ringuet, \'Ecole polytechnique, CNRS/IN2P3, Palaiseau, France \\ 
$^2$Institute of High Energy Physics, Beijing, China \\ 
$^3$Department of Physics, University of Tokyo, Japan}


\date{\today}

\begin{abstract}

We explore the fractal nature of particle showers using Monte-Carlo simulation. 
We define the fractal dimension of showers measured in a high granularity calorimeter designed for a future lepton collider.
The shower fractal dimension reveals detailed information of the spatial configuration of the shower.
It is found to be characteristic of the type of interaction and highly sensitive to the nature of the incident particle. 
Using the shower fractal dimension, we demonstrate a particle identification algorithm that can efficiently separate electromagnetic showers, hadronic showers and non-showering tracks. 
We also find a logarithmic dependence of the shower fractal dimension on the particle energy. 
\end{abstract}


\maketitle

\emph{Introduction.---}
When an energetic particle impinges on matter, it may interact and produce daughter particles, which may themselves interact. 
This process iterates while daughter particles have sufficient energy. 
The resulting particle cascade is called a shower~\cite{Cassen:1933sf, Auger:1938ef}. 
A profound understanding of particle showers, a fundamental phenomenon of particle - matter interactions, is crucial for experimental high energy physics, astrophysics, radiation protection and radiotherapy.

\begin{figure}
\centerline{\includegraphics[width=0.98\columnwidth]{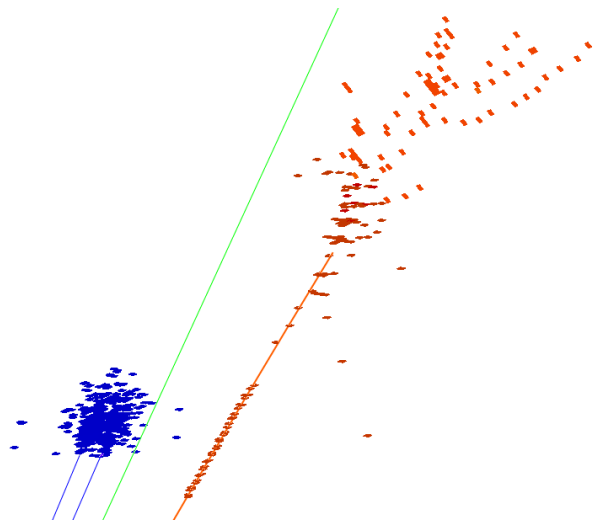}}
\caption{ A simulated $\tau^{+}\rightarrow2\gamma(\pi^{0})+\pi^{+}+\bar{\nu}_{\tau}$ event in a linear collider calorimeter. Photons and their showers (electromagnetic) are colored blue, $\pi^{+}$ and its shower (hadronic) are colored red. The green line indicates the neutrino trajectory, which roughly corresponding to the direction of $\tau^{+}$. The detector hits are displayed according to their size ($10 \times{10}~{mm}^{2}$) and orientation.}
\label{Showersample}
\end{figure}

Showers can be classified into electromagnetic and hadronic types. 
The development of electromagnetic showers is governed by $e^{+}e^{-}$ pair-production and bremsstrahlung interactions.
Hadronic showers are composed of long hadron tracks and localised clusters produced in $\pi^{0}$ decay ($\pi^{0}\rightarrow\gamma\gamma$) or nuclear breakup~\cite{Wigmans}. 
The strong interactions between nuclei and hadrons, particularly pion generation, determine the development of hadronic showers.
The typical spatial configurations of these two showers types are illustrated in Fig.~\ref{Showersample}.

These cascade mechanisms give rise to the fractal nature of particle showers.
The fractal structure of high energy cosmic showers in the atmosphere has been previously studied~\cite{Ohnishi:1990tv, Haungs:1995dn, Kempa:1998gb}. 
In this letter, we explore for the first time the fractal nature of particle showers produced and measured in a calorimeter. 
This calorimeter is designed for high energy physics experiments with ultra-high granularity.
We observe a strong dependence of the number of hits obtained when the effective granularity of the calorimeter readout is varied, from which we define the shower fractal dimension.
We investigate the dependence of the shower fractal dimension on the type and energy of the incident particle, and demonstrate a particle identification algorithm based only on measurements made with the calorimeter. 

\emph{Method and Measurement.---}
The detector used in this study is a hadron calorimeter designed for a future $e^{+}e^{-}$ linear collider~\cite{CLICCDR, LCRDR}.
The calorimeter structure follows the geometry of the barrel hadron calorimeter (HCAL) of the International Large Detector~\cite{LoISS}. 
Prototypes of such an HCAL have been developed by the CALICE collaboration~\cite{DHCAL, SDHCAL}. 
The calorimeter consists of 48 layers of 20~mm thick iron absorbers, interleaved with 6.5~mm thick resistive plate chambers (RPC).
In the CALICE prototypes, the RPCs are read out in binary mode with a granularity of $10 \times{10}~{mm}^{2}$.
Such high granularity is required by particle flow algorithms, which can achieve excellent jet energy resolution by separating the individual particles in a jet and measuring them in the most suited subdetectors.
It also provides a detailed view of the showers forgotten since the age of heavy liquid bubble chambers~\cite{Gargamelle}.

The typical lateral dispersion of induced charge in the RPC is around 1~mm, 
so $1 \times{1}~{mm}^{2}$ represents the ultimate granularity achievable with this technology. 
In this study, we simulated a $1 \times{1}~{mm}^{2}$ readout granularity.
The interactions of different particles ($\pi^{+}$, $\mu^{+}$, $e^{+}$, $K^{0}_{L}$) in the calorimeter were simulated. 

The effective readout cell size can be varied by grouping blocks of $\alpha \times{\alpha}$ cells, where $\alpha$ defining the scale at which the shower is analysed.
Defining $N_{\alpha}$ as the number of hits at scale $\alpha$, the ratio of the number of hits at different scales can be written as:
\begin{eqnarray*}
\centering
R_{\alpha, \beta} = N_{\beta}/N_{\alpha}.
\label{RatioDef}
\end{eqnarray*}
Choosing $\beta$ to be smaller than $\alpha$, this ratio is then equal to or larger than 1.
To make the best use of the recorded spatial information, $\beta$ can be set to 1, corresponding to the ultimate cell size.
In our study, $\beta$ has been set to be either 1 or 10.
The former is used to explore the self-similar behavior of showers at a small scale, while the latter is used to estimate the performance at a scale realized in current calorimeter prototypes~\cite{DHCAL, SDHCAL}. 

\begin{figure}
\centerline{\includegraphics[width=1.0\columnwidth]{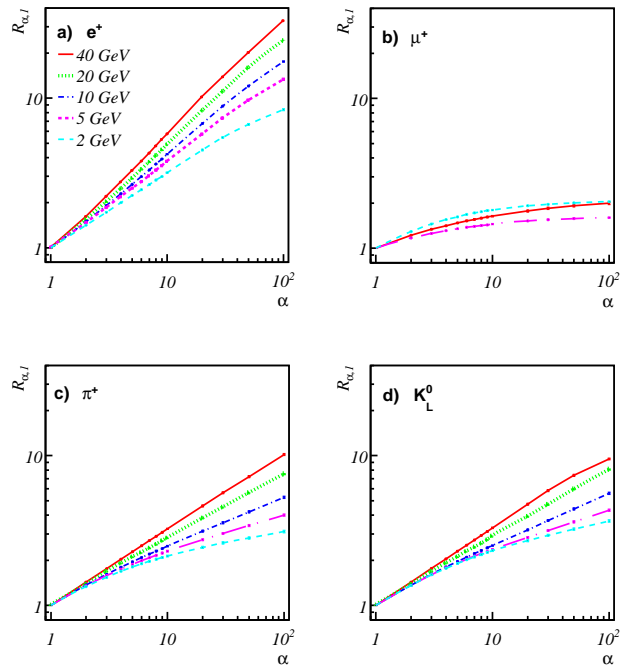}}
\caption{Dependence of the ratio $R_{\alpha, 1}$ versus the scale $\alpha$ for different samples, 1k events were simulated per sample per energy point. (a) $e^{+}$, (b) $\mu^{+}$, (c) $\pi^{+}$ and (d) $K_{L}^{0}$.}
\label{AllSample}
\end{figure}

Fig.~\ref{AllSample} shows the correlation between the hit ratio $R_{\alpha, 1}$ and the scale $\alpha$ for various samples.
For both electromagnetic and hadronic showers, a linear dependence is observed on a double logarithmic scale. 
The shower fractal dimension can therefore be defined as 

\begin{equation}
\centering
FD_{\beta} = \left< \frac{\log(R_{\alpha, \beta})}{\log(\alpha)} \right> + 1. 
\label{SFDdef}
\end{equation}
The first term represents the average slope of the correlation shown in Fig.~\ref{AllSample}, while the second term is due to the longitudinal degree of freedom, since the effective cell size is varied only within detector layers.
With increasing scale, the number of hits converges to the number of fired layers, and these curves therefore saturate at large scale. 
An adequate scale range is needed to calculate the fractal dimension.
In this analysis, 16 ratios were used to calculate $FD_{1 mm}$: $R_{i, 1}$ (i = 2-10, 20, 30, 50, 60, 90, 120, 150), 
while 7 ratios were used at $\beta = 10$: $R_{i, 10}$ (i = 20, 30, 50, 60, 90, 120, 150).
Muons usually induce only a non-showering track in the calorimeter, which can be regarded as an extreme case of a particle shower.
Their fractal dimension is thus also measured using the same method (Fig.~\ref{AllSample}(b)).

The main features of the curves shown in Fig.~\ref{AllSample} can be understood qualitatively.
In the ideal case, non-showering particles such as $\mu^{+}$ deposit only one hit per layer.
Therefore the number of hits is almost insensitive to the scale.
The first term in Eq.~\ref{SFDdef} vanishes, yielding a fractal dimension of 1. 
Electromagnetic showers are the most compact and exhibit the largest fractal dimension. 
As discussed above, hadronic showers are composed of tracks from charged hadrons and localized electromagnetic sub-showers, giving fractal dimensions with a value between those of electromagnetic showers and non-showering particles. 
The fractal dimensions of different hadrons ($\pi^{+}$, $K_{L}^{0}$) at the same energy are very similar.
Similar curves were observed for neutrons and protons.
Because $\pi^{+}$ showers have minimum ionizing particle tracks before the first interaction while $K_{L}^{0}$ showers do not, the mean fractal dimension of $\pi^{+}$ showers is slightly smaller than that of $K_{L}^{0}$ showers at a given energy. 

\begin{figure}
		\centerline{\includegraphics[width=1.0\columnwidth]{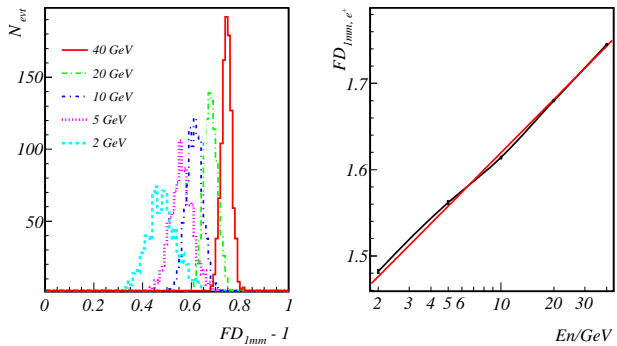}}
		\caption{ Left: Fractal dimension of positron samples at different energies. Right: Correlation between fractal dimension and particle energy. }
		\label{ele_FD_E}
		\end{figure}

Fig.~\ref{ele_FD_E} shows the measured fractal dimension for $e^{+}$ samples at different energies, where a linear dependence of the average FD on the logarithm of the incident particle energy is observed. 
As shown in Fig.~\ref{ele_FD_E}, the average shower fractal dimension scales approximately as: 
\begin{eqnarray}
\displaystyle FD^{\mu}_{1 \mathrm{mm}}(E) &=& 1.2       \nonumber \\
\displaystyle FD^{em}_{1 \mathrm{mm}}(E) &=& 1.41 + 0.21 \times{\log_{10}(E/\mathrm{GeV})}  \\
\displaystyle FD^{had}_{1 \mathrm{mm}}(E) &=& 1.24 + 0.15 \times{\log_{10}(E/\mathrm{GeV})}	\nonumber
\label{MIPFD}
\end{eqnarray}
where E is the incident particle energy.

		\begin{figure}
		\centerline{\includegraphics[width=0.99\columnwidth]{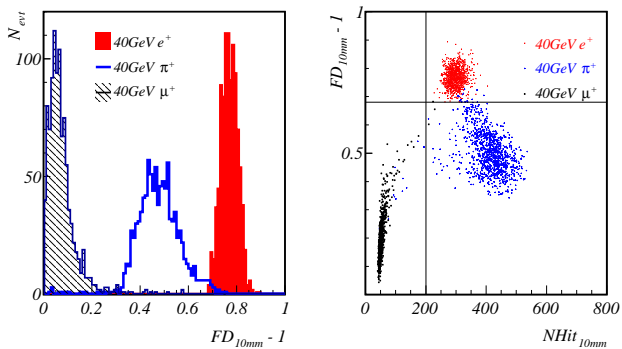}}
		\caption{ 40GeV $\mu^{+}$, $e^{+}$ and $\pi^{+}$ samples at 10~mm calorimeter cell size. Left: Fractal dimension, Right: Fractal dimension versus number of hits }
		\label{FD_1d}
		\end{figure}

		\emph{Application to particle identification.---}
		To be applicable to current HCAL prototype data, the following discussion is based on the fractal dimension using $10 \times{10}~{mm}^{2}$ readout cells.
		Mathematically, the fractal dimension is rigorously defined at infinitesimally small sizes, but in practice, as long as the particle shower creates more than 10 hits, the measured fractal dimension can reflect the nature of the incident particle.
		Since each hadronic shower hit in the prototypes is roughly equivalent to 100~MeV energy deposition~\cite{DHCAL, SDHCAL}, we can measure the fractal dimension of showers with energy of at least 1~GeV with the current prototypes.
	
		The distributions of $FD_{10 mm}$ of 40~GeV $\mu^{+}$, $e^{+}$ and $\pi^{+}$ showers are shown in the left plot of Fig.~\ref{FD_1d}.
		These distributions are well separated. 
		The $FD_{10 mm}$ distribution of the $e^{+}$ sample is approximately Gaussian, while that of the $\mu^{+}$ sample is peaked at low values with a positive tail due to bremsstrahlung photons.
		The distribution of the $\pi^{+}$ sample lies between the other two with slight overlaps.  
		These overlaps are mainly caused by experimentally indistinguishable events. 
		A $\pi^{+}$ shower can have a fractal dimension comparable to a $\mu^{+}$ shower in the case of pion decay ($\pi^{+} \rightarrow \mu^{+} + \nu_{\mu}$) before reaching the calorimeter. 
		On the contrary, if the majority of the $\pi^{+}$ energy is deposited electromagnetically, the fractal dimension of the $\pi^{+}$ shower can be close to that of $e^{+}$ showers. 
		For example, a charged pion may convert into $\pi^{0}$ through isospin exchange ($\pi^{+} + n \rightarrow \pi^{0} + p$).

		The separation in Fig.~\ref{FD_1d} indicates that the measured shower fractal dimension can be used for particle identification, a key task of event reconstruction in experimental high energy physics. 
		The right plot of Fig.~\ref{FD_1d} shows the distribution of shower fractal dimension versus number of hits for 40~GeV $\mu^{+}$, $e^{+}$ and $\pi^{+}$ samples.
		The fact that these samples are clearly separated suggests that even straightforward cuts can provide efficient particle identification.
		Showers with $FD \geqslant 0.68$ are identified as electromagnetic, those with $FD < 0.68$ and $N_{hits} > 200$ as hadronic, and the rest as $\mu^{+}$ (as indicated by the black lines in the right plot of Fig.~\ref{FD_1d}).
		The performance of this selection is shown in Table~\ref{PIDtable}. 

		\begin{table}[h]
		\caption{\label{PIDtable} Performance of particle identification based on shower fractal dimension, the particle energy is fixed at 40~GeV.}
		\begin{center}
		\begin{ruledtabular}
		\begin{tabular}{*{4}{l}}
		& $e^{+}$ & $\mu^{+}$ & $\pi^{+}$  \\
			$e^{+}$   & 100\%    & 0         & 0			\\
			$\mu^{+}$ & 0       & 99.5\%       & 0.5\%			\\
			$\pi^{+}$ & 1.7\%      & 1.4\%        & 96.9\%		\\
			\end{tabular}
			\end{ruledtabular}
			\end{center}
			\end{table}

\begin{figure}
			\centerline{\includegraphics[width=0.99\columnwidth]{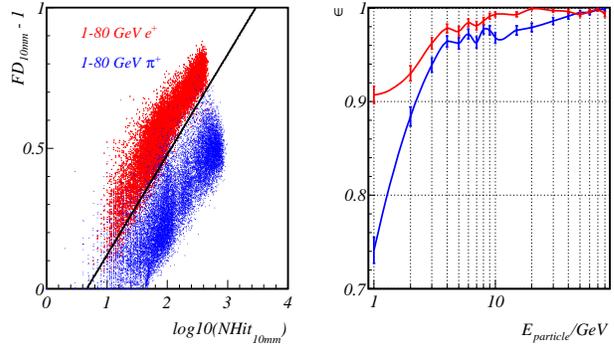}}
			\caption{ Left: Distribution of 1-80~GeV $e^{+}$ and $\pi^{+}$ samples in the plane of fractal dimension versus number of hits. Right: Efficiency of tagging electromagnetic (red curve) and hadronic (blue curve) showers at different energies with the cut indicated by the black line in the left plot.}
			\label{FD_1d_FullRange}
			\end{figure}

			The distribution of the $e^{+}$ and $\pi^{+}$ samples in the plane of these two observables depends on the initial particle energy. 
			Fig.~\ref{FD_1d_FullRange} shows the distributions of $e^{+}$ and $\pi^{+}$ samples over an energy range of 1 to 80~GeV. 
			For both types, samples are simulated at 18 energy points (1, $2\rightarrow10$~GeV, 15~GeV, 20~GeV, $30\rightarrow80$~GeV) with 1000 events at each point.  
			Both samples exhibit a linear dependence of the shower fractal dimension on the logarithm of the number of hits, supporting the observation that the shower fractal dimension is logarithmically dependent on the particle energy. 

			The $e^{+}$ and $\pi^{+}$ samples are well separated over the entire energy range. 
			This global separation can be used to identify the type of incident particle without reference to its energy.
			For example, a global cut of 0.67 on the combined observable $log_{10}(N_{hits}) - 2.8\times{FD_{10 mm}}$ (indicated by the black line in the left plot of Fig.~\ref{FD_1d_FullRange}) can correctly identify 98\% of electromagnetic and 96\% of hadronic showers, averaged over the whole energy range.
			The energy dependence of these efficiencies is shown in the right plot of Fig.~\ref{FD_1d_FullRange}. 
			The inefficiency is mainly due to fluctuations at low energies.
			If we restrict the sample energy to be higher than 10~GeV, the efficiency of electromagnetic and hadronic shower tagging can reach 99.5\% and 98.7\% respectively.

			The particle identification algorithm demonstrated above is different from conventional methods in two aspects:
			first, the shower fractal dimension is based solely on calorimeter data, which avoids possible bias when combining measurements from different detectors (for example, one of the most discriminating variables, the $E/P$ ratio, requires both the particle energy measured with the calorimeter and the particle momentum measured with the tracker).
			Secondly, the shower fractal dimension is defined in transverse directions, and is therefore orthogonal to the longitudinal shower profiles widely used in conventional particle identification algorithms.
			The performance of this fractal-dimension-based algorithm is comparable to that of the $Z\rightarrow\tau\tau$ particle identification result of ALEPH~\cite{ALEPHPID}, which makes use of tracker, calorimeter and Muon chambers. 
			Of course, other measurements can be combined to achieve a better result, however, the performance of this method is already close to the limit where the inefficiency is dominated by indistinguishable processes of charged $\pi$ decay and isospin exchange.
			This method is also valid for particles with energy as low as 1~GeV, where particle identification with the conventional calorimeter measurements becomes difficult~\cite{DHCAL}. 

			\emph{Conclusions.---}
			Using a sampling calorimeter with high transverse granularity (1 $cm^{2}$ cell size), we explore the expected transverse self-similar pattern of particle showers.
			We observe a clear logarithmic dependence of the number of hit cells on the cell size over an adequate range of scales (from 1~mm upto 100~mm),
			from which we define the shower fractal dimension.
			The shower fractal dimension reveals detailed information of showers' spatial configuration, and is found to be characteristic of the nature of the impinging particle.
			Using the fractal dimension, we demonstrate a particle identification algorithm based purely on calorimeter observables, which can distinguish electromagnetic showers, hadronic showers and non-showering tracks with efficiencies close to the physical limit, and valid for particles with energy as low as 1~GeV.
			A logarithmic dependence between the shower fractal dimension and impinging particle energy is also observed.	

			The author thank Y.~Liu and L.~W.~Yu for their support and disucssion. This project has been partially funded by the European Commission within Framework Program 7 Capacities, Grant Agreement No 262025.

			\bibliographystyle{apsrev}

			\bibliography{FD.bib}

\begin{thebibliography}{13}
\expandafter\ifx\csname natexlab\endcsname\relax\def\natexlab#1{#1}\fi
\expandafter\ifx\csname bibnamefont\endcsname\relax
  \def\bibnamefont#1{#1}\fi
\expandafter\ifx\csname bibfnamefont\endcsname\relax
  \def\bibfnamefont#1{#1}\fi
\expandafter\ifx\csname citenamefont\endcsname\relax
  \def\citenamefont#1{#1}\fi
\expandafter\ifx\csname url\endcsname\relax
  \def\url#1{\texttt{#1}}\fi
\expandafter\ifx\csname urlprefix\endcsname\relax\def\urlprefix{URL }\fi
\providecommand{\bibinfo}[2]{#2}
\providecommand{\eprint}[2][]{\url{#2}}

\bibitem[{\citenamefont{Cassen}(1933)}]{Cassen:1933sf}
\bibinfo{author}{\bibfnamefont{B.}~\bibnamefont{Cassen}},
  \bibinfo{journal}{Phys.Rev.} \textbf{\bibinfo{volume}{44}},
  \bibinfo{pages}{513} (\bibinfo{year}{1933}).

\bibitem[{\citenamefont{Auger et~al.}(1938)\citenamefont{Auger, Maze, and
  Grivet-Mayer}}]{Auger:1938ef}
\bibinfo{author}{\bibfnamefont{P.}~\bibnamefont{Auger}},
  \bibinfo{author}{\bibfnamefont{R.}~\bibnamefont{Maze}}, \bibnamefont{and}
  \bibinfo{author}{\bibfnamefont{T.}~\bibnamefont{Grivet-Mayer}},
  \bibinfo{journal}{C.R.Acad.Sci.} \textbf{\bibinfo{volume}{206}},
  \bibinfo{pages}{1721} (\bibinfo{year}{1938}).

\bibitem[{\citenamefont{Wigmans}(2000)}]{Wigmans}
\bibinfo{author}{\bibfnamefont{R.}~\bibnamefont{Wigmans}},
  \emph{\bibinfo{title}{Calorimetry Energy Measurement in Particle Physics}}
  (\bibinfo{publisher}{Clarendon Press}, \bibinfo{year}{2000}).

\bibitem[{\citenamefont{Ohnishi}(1990)}]{Ohnishi:1990tv}
\bibinfo{author}{\bibfnamefont{T.}~\bibnamefont{Ohnishi}},
  \bibinfo{journal}{Can.J.Phys.} \textbf{\bibinfo{volume}{68}},
  \bibinfo{pages}{906} (\bibinfo{year}{1990}).

\bibitem[{\citenamefont{Haungs et~al.}(1996)\citenamefont{Haungs, Kempa,
  Mathes, Rebel, and Wentz}}]{Haungs:1995dn}
\bibinfo{author}{\bibfnamefont{A.}~\bibnamefont{Haungs}},
  \bibinfo{author}{\bibfnamefont{J.}~\bibnamefont{Kempa}},
  \bibinfo{author}{\bibfnamefont{H.}~\bibnamefont{Mathes}},
  \bibinfo{author}{\bibfnamefont{H.}~\bibnamefont{Rebel}}, \bibnamefont{and}
  \bibinfo{author}{\bibfnamefont{J.}~\bibnamefont{Wentz}},
  \bibinfo{journal}{Nucl.Instrum.Meth.} \textbf{\bibinfo{volume}{A372}},
  \bibinfo{pages}{515} (\bibinfo{year}{1996}).

\bibitem[{\citenamefont{Kempa and Samorski}(1998)}]{Kempa:1998gb}
\bibinfo{author}{\bibfnamefont{J.}~\bibnamefont{Kempa}} \bibnamefont{and}
  \bibinfo{author}{\bibfnamefont{M.}~\bibnamefont{Samorski}},
  \bibinfo{journal}{J.Phys.} \textbf{\bibinfo{volume}{G24}},
  \bibinfo{pages}{1039} (\bibinfo{year}{1998}).

\bibitem[{\citenamefont{Aicheler et~al.}(2012)\citenamefont{Aicheler, Burrows,
  Draper, Garvey, Lebrun, Peach, Phinney, Schmickler, Schulte, and
  Toge}}]{CLICCDR}
\bibinfo{author}{\bibfnamefont{M.}~\bibnamefont{Aicheler}},
  \bibinfo{author}{\bibfnamefont{P.}~\bibnamefont{Burrows}},
  \bibinfo{author}{\bibfnamefont{M.}~\bibnamefont{Draper}},
  \bibinfo{author}{\bibfnamefont{T.}~\bibnamefont{Garvey}},
  \bibinfo{author}{\bibfnamefont{P.}~\bibnamefont{Lebrun}},
  \bibinfo{author}{\bibfnamefont{K.}~\bibnamefont{Peach}},
  \bibinfo{author}{\bibfnamefont{N.}~\bibnamefont{Phinney}},
  \bibinfo{author}{\bibfnamefont{H.}~\bibnamefont{Schmickler}},
  \bibinfo{author}{\bibfnamefont{D.}~\bibnamefont{Schulte}}, \bibnamefont{and}
  \bibinfo{author}{\bibfnamefont{N.}~\bibnamefont{Toge}}, \bibinfo{type}{Tech.
  Rep.} \bibinfo{number}{CERN-2012-007. SLAC-R-985. KEK-Report-2012-1.
  PSI-12-01. JAI-2012-001}, \bibinfo{institution}{CERN},
  \bibinfo{address}{Geneva} (\bibinfo{year}{2012}).

\bibitem[{\citenamefont{Brau et~al.}(2007)\citenamefont{Brau, Okada, Walker,
  Djouadi, Lykken, Monig, Oreglia, Yamashita, Phinney, Toge et~al.}}]{LCRDR}
\bibinfo{author}{\bibfnamefont{J.}~\bibnamefont{Brau}},
  \bibinfo{author}{\bibfnamefont{Y.}~\bibnamefont{Okada}},
  \bibinfo{author}{\bibfnamefont{N.~J.} \bibnamefont{Walker}},
  \bibinfo{author}{\bibfnamefont{A.}~\bibnamefont{Djouadi}},
  \bibinfo{author}{\bibfnamefont{J.}~\bibnamefont{Lykken}},
  \bibinfo{author}{\bibfnamefont{K.}~\bibnamefont{Monig}},
  \bibinfo{author}{\bibfnamefont{M.}~\bibnamefont{Oreglia}},
  \bibinfo{author}{\bibfnamefont{S.}~\bibnamefont{Yamashita}},
  \bibinfo{author}{\bibfnamefont{N.}~\bibnamefont{Phinney}},
  \bibinfo{author}{\bibfnamefont{N.}~\bibnamefont{Toge}}, \bibnamefont{et~al.},
  \emph{\bibinfo{title}{International Linear Collider reference design report:
  ILC Global Design Effort and World Wide Study}} (\bibinfo{publisher}{CERN},
  \bibinfo{address}{Geneva}, \bibinfo{year}{2007}).

\bibitem[{\citenamefont{Abe et~al.}(2010)}]{LoISS}
\bibinfo{author}{\bibfnamefont{T.}~\bibnamefont{Abe}} \bibnamefont{et~al.},
  \bibinfo{type}{Letter of intent} \bibinfo{number}{DESY-2009-87},
  \bibinfo{institution}{DESY} (\bibinfo{year}{2010}).

\bibitem[{\citenamefont{Bilki et~al.}(2009)\citenamefont{Bilki, Butler,
  Mavromanolakis, May, Norbeck, Repond, Underwood, Xia, and Zhang}}]{DHCAL}
\bibinfo{author}{\bibfnamefont{B.}~\bibnamefont{Bilki}},
  \bibinfo{author}{\bibfnamefont{J.}~\bibnamefont{Butler}},
  \bibinfo{author}{\bibfnamefont{G.}~\bibnamefont{Mavromanolakis}},
  \bibinfo{author}{\bibfnamefont{E.}~\bibnamefont{May}},
  \bibinfo{author}{\bibfnamefont{E.}~\bibnamefont{Norbeck}},
  \bibinfo{author}{\bibfnamefont{J.}~\bibnamefont{Repond}},
  \bibinfo{author}{\bibfnamefont{D.}~\bibnamefont{Underwood}},
  \bibinfo{author}{\bibfnamefont{L.}~\bibnamefont{Xia}}, \bibnamefont{and}
  \bibinfo{author}{\bibfnamefont{Q.}~\bibnamefont{Zhang}},
  \bibinfo{journal}{JINST} \textbf{\bibinfo{volume}{4}},
  \bibinfo{pages}{P10008} (\bibinfo{year}{2009}).

\bibitem[{\citenamefont{Bedjidian et~al.}(2011)\citenamefont{Bedjidian,
  Belkadhi, Boudry, Combaret, Decotigny, Gil, de~la Taille, Dellanegra,
  Gapienko, Grenier et~al.}}]{SDHCAL}
\bibinfo{author}{\bibfnamefont{M.}~\bibnamefont{Bedjidian}},
  \bibinfo{author}{\bibfnamefont{K.}~\bibnamefont{Belkadhi}},
  \bibinfo{author}{\bibfnamefont{V.}~\bibnamefont{Boudry}},
  \bibinfo{author}{\bibfnamefont{C.}~\bibnamefont{Combaret}},
  \bibinfo{author}{\bibfnamefont{D.}~\bibnamefont{Decotigny}},
  \bibinfo{author}{\bibfnamefont{E.~C.} \bibnamefont{Gil}},
  \bibinfo{author}{\bibfnamefont{C.}~\bibnamefont{de~la Taille}},
  \bibinfo{author}{\bibfnamefont{R.}~\bibnamefont{Dellanegra}},
  \bibinfo{author}{\bibfnamefont{V.~A.} \bibnamefont{Gapienko}},
  \bibinfo{author}{\bibfnamefont{G.}~\bibnamefont{Grenier}},
  \bibnamefont{et~al.}, \bibinfo{journal}{JINST} \textbf{\bibinfo{volume}{6}},
  \bibinfo{pages}{P02001} (\bibinfo{year}{2011}).

\bibitem[{\citenamefont{Behr and Mittner}(1963)}]{Gargamelle}
\bibinfo{author}{\bibfnamefont{L.}~\bibnamefont{Behr}} \bibnamefont{and}
  \bibinfo{author}{\bibfnamefont{P.}~\bibnamefont{Mittner}},
  \bibinfo{journal}{Nucl.Instrum.Meth.} \textbf{\bibinfo{volume}{20}},
  \bibinfo{pages}{446} (\bibinfo{year}{1963}).

\bibitem[{\citenamefont{Heister et~al.}(2001)}]{ALEPHPID}
\bibinfo{author}{\bibfnamefont{A.}~\bibnamefont{Heister}} \bibnamefont{et~al.}
  (\bibinfo{collaboration}{ALEPH Collaboration}),
  \bibinfo{journal}{Eur.Phys.J.} \textbf{\bibinfo{volume}{C20}},
  \bibinfo{pages}{401} (\bibinfo{year}{2001}), \eprint{hep-ex/0104038}.

\end{thebibliography}

			\end{document}